\documentclass[epj]{webofc}
\usepackage[varg]{txfonts}   
\usepackage{amsmath}
\usepackage{epsfig}
\usepackage{graphicx}
\usepackage{slashed}
\usepackage{color}

\def\bea{\begin{eqnarray}}
\def\eea{\end{eqnarray}}
\def\be{\begin{equation}}
\def\ee{\end{equation}}

\def\lQ{\Lambda_{\rm QCD}}

\def\als{\alpha_{\rm s}}

\def\siml{{\ \lower-1.2pt\vbox{\hbox{\rlap{$<$}\lower6pt\vbox{\hbox{$\sim$}}}}\ }}     
\def\simg{{\ \lower-1.2pt\vbox{\hbox{\rlap{$>$}\lower6pt\vbox{\hbox{$\sim$}}}}\ }} 

\newcommand{\MS}{\overline{\rm MS}}
\definecolor{bblue}{rgb}{0,0,0.7}
\definecolor{bglight}{rgb}{1,1,1}
\definecolor{white}{rgb}{1,1,1}
\definecolor{WHITE}{rgb}{1,1,1}
\definecolor{lyellow}{rgb}{0.9,0.9,0.8} 
\definecolor{lgreen}{rgb}{1,0.5,1}
\definecolor{dgreen}{rgb}{0,0.5,0.5}
\definecolor{GGreen}{rgb}{0,1,0}
\definecolor{Green}{rgb}{0,0.75,0}
\definecolor{lightGreen}{rgb}{0.85,1,0.75}
\definecolor{dblue}{rgb}{0,0,0.6}
\definecolor{lblue}{rgb}{0.9,0.9,1}
\definecolor{lb}{rgb}{0.5,0.5,1}
\definecolor{grey}{rgb}{0.3,0.3,0.3}
\definecolor{lgrey}{rgb}{0.92,0.92,0.92}
\definecolor{llgrey}{rgb}{0.6,0.6,0.6}
\definecolor{Black}{rgb}{0,0,0}
\definecolor{LemonChiffon}{rgb}{1.,0.98,0.8}
\definecolor{TUMblue}{rgb}{0.157,0.451,0.706}
\definecolor{celeste}{rgb}{0.9,0.95,1}
\definecolor{pink}{rgb}{1,0.1,1}
\definecolor{yellow}{rgb}{0.5,0.5,0}
\definecolor{orange}{rgb}{0.75,0.25,0}

\woctitle{XLV International Symposium on Multiparticle Dynamics}

\begin{document}

\title{A low-energy determination of $\als$ at three loops}

\author{Antonio Vairo \inst{1}\fnsep\thanks{\email{antonio.vairo@ph.tum.de}}}

\institute{Physik-Department, Technische Universit\"{a}t M\"{u}nchen, \\James-Franck-Str. 1, 85748 Garching, Germany}

\abstract{ 
We review one of the most accurate low-energy determinations of $\als$.
Comparing at short distances the QCD static energy at three loops and resummation of the next-to-next-to leading logarithms 
with its determination in 2+1-flavor lattice QCD, 
we obtain $\als(1.5~{\rm GeV})=0.336^{+0.012}_{-0.008}$, which corresponds to
$\als(M_Z)=0.1166^{+0.0012}_{-0.0008}$. We discuss future perspectives.
}

\maketitle

\section{$\als$ in 2015}
\label{introduction}
\begin{figure}[hb]
\centering
\includegraphics[width=7.5cm]{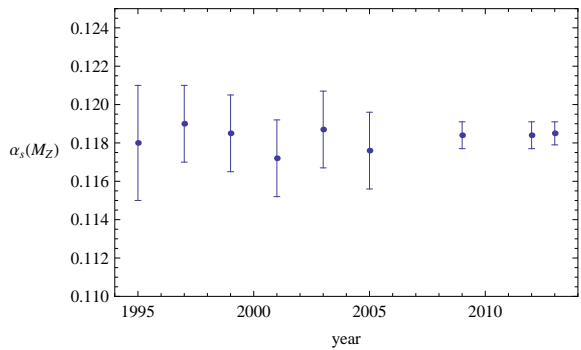}
\caption{List of PDG averages for $\als(M_Z)$ from 1995 to 2013, source~\cite{PDGarchive}.}
\label{figPDG}
\end{figure}

For many years the average of the strong coupling constant, $\als$, 
provided by the Particle Data Group (PDG) (the latest printed edition is~\cite{Agashe:2014kda})
has shown a rather stable central value and a steady decrease in the associated error, see figure~\ref{figPDG}.
This satisfactory situation has however been challenged in the last years (see for example the summary tables 
in~\cite{Moch:2014tta,Bazavov:2014soa,Hoang:2015hka} and the recent collection~\cite{d'Enterria:2015toz}): 
precise determinations pointing often towards a lower value of $\als$ have appeared, 
traditionally larger and precise determinations from $\tau$ decay have become smaller and less precise~\cite{Boito:2014sta}, 
accurate determinations from lattice QCD have been questioned.
The 2016 PDG average will reflect this changed situation and, for the first time in twenty years, 
the $\als$ world average will see an increase in the error~\cite{Bethke2015}.

This change of scenario makes it even more important to look for accurate determinations of $\als$.
The extraction of $\als$ that follows from comparing perturbative QCD with short distance lattice computations of the QCD static 
energy is one of these determinations. Perturbatively this quantity is known at three loops and 
next-to-next-to-next-to leading logarithmic (NNNLL) accuracy. On the lattice side, the static Wilson loop 
is one of the most accurately known quantities that has been computed since the establishment 
of lattice QCD. Nowadays computations are with quarks at or very close to their physical masses
so that the lattice spacing can be matched to a physical scale, eventually leading to  
a determination of the physical (unquenched) $\als$. The strong coupling constant extracted in this way is 
typically at a low energy scale because the lattice cannot explore too short distances. 
In fact the method provides one of the most precise low-energy determinations of 
$\als$. In this sense the determination of $\als$ from the QCD static energy not only provides 
a competitive but also a complementary determination with respect to high-energy determinations.
Obviously there is also a value {\it per se} in having a low-energy determination of $\als$, 
for it adds to our understanding of low-energy QCD and non-trivially constraints our modeling of it.
Here we will review our latest determination of $\als$ from the QCD static energy~\cite{Bazavov:2014soa},
which improves our previous determination~\cite{Bazavov:2012ka}. An earlier quenched analysis can be found in~\cite{Brambilla:2010pp}. 
Finally, as we will discuss briefly in the outlook, although the extraction of $\als$ from the static 
Wilson loop provides already today an accurate value, there is room for improvement in several directions.

\section{QCD static energy in perturbation theory}
\label{theory}
The QCD static energy, $E_0(r)$, is defined (in Minkowski spacetime) as
\be
E_0(r) = \lim_{ T\to\infty}\frac{i}{T} \ln \, \left\langle {\rm Tr} \,{\rm P} \exp\left\{i g \oint_{r\times T} dz^\mu \, A_\mu\right\} \right\rangle, 
\ee
where the integral is over a rectangle of spatial length $r$, the distance between the static quark and the static 
antiquark, and time length $T$; $\langle \dots \rangle$ stands for the path integral over the gauge fields $A_\mu$ and 
the light quark fields, P is the path-ordering operator of the color matrices and $g$ is the SU(3) gauge coupling ($\als = g^2/(4\pi)$). 
The above definition of $E_0(r)$ is valid at any distance $r$.
In the short range, $r\lQ \ll 1$ for which $\als(1/r) \ll 1$, $E_0(r)$ may be computed in perturbative QCD 
and expressed as a series in $\als$ (computed at a typical scale of order $1/r$):
\be
E_0(r) = \Lambda_s - \frac{4\als}{3r}(1 + \dots),
\ee
where $\Lambda_s$ is a constant that accounts for the normalization of the static energy and the dots stand for higher-order terms.
The expansion of $E_0(r)$ in powers of $\als$ is known up to three loops.
At three loops, a contribution proportional to $\ln \als$ appears for the first time.
This three-loop logarithm has been computed in~\cite{Brambilla:1999qa,Brambilla:1999xf}.
The complete three-loop contribution can be derived from~\cite{Anzai:2009tm,Smirnov:2009fh}.
The leading logarithms have been resummed to all orders in~\cite{Pineda:2000gza}, 
providing, among others, also the four-loop contribution proportional to $\ln^2 \als$.
The four-loop contribution proportional to $\ln \als$ has been computed in~\cite{Brambilla:2006wp}.
Next-to-leading logarithms have been resummed to all orders in~\cite{Brambilla:2009bi}, 
which represents the status of the art of the QCD static energy in perturbation theory.
A concise but complete summary with all relevant formulas can be found in~\cite{Tormo:2013tha}. 
$E_0(r)$ is, therefore, one of the best known quantities in perturbative QCD.

\begin{figure}[ht]
\centering
\includegraphics[width=7.5cm]{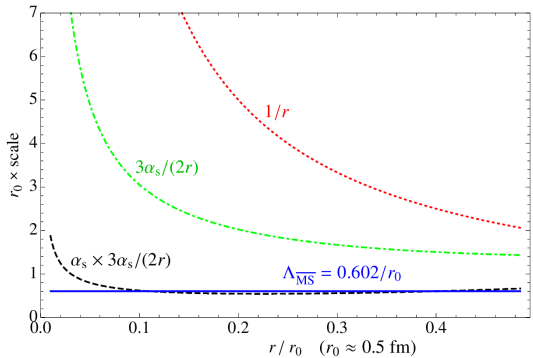}
\caption{Energy scales for the short-range static energy in the quenched case, from~\cite{Brambilla:2009bi}.}
\label{figscales}
\end{figure}

The appearance of $\ln \als$ terms starting from three loops signals the cancellation of contributions 
coming from the different energy scales $1/r$ and $\als/r$:
\be
\ln \als = \ln \frac{\mu}{1/r} + \ln\frac{\als/r}{\mu}, 
\ee
where $\mu$ is a factorization scale. Indeed in the short range $E_0(r)$ is characterized by a hierarchy 
of well-separated scales, see figure~\ref{figscales}.

The separation of contributions coming from gluons scaling like $1/r$ from contributions coming from gluons scaling like $\als/r$ 
leads to the factorization formula~\cite{Brambilla:1999qa,Brambilla:1999xf}:
\bea
E_0(r) &=& \Lambda_s ~~~ + ~~~V_{s}(r, \mu) - i \frac{V_A^2}{3} \int_0^\infty dt \, e^{-i t (V_o-V_s)}
\,\langle {\rm Tr} \,\{ {\bf r} \cdot g{\bf E}(t,{\bf 0}) \, {\bf r} \cdot g{\bf E}(0,{\bf 0})\}  \rangle(\mu) + \dots \;,
\label{E0}
\eea
where $V_s(r,\mu) = -4\als/(3r) + \dots$ is the color-singlet static potential, 
$V_o(r,\mu) = \als/(6r) + \dots$ is the color-octet static potential, $V_A$ a Wilson coefficient describing the 
chromoelectric dipole coupling, and ${\bf E}$ the chromoelectric field. The color-singlet static 
potential encodes the contributions from the scale $1/r$, while the low-energy contributions are in the 
term proportional to the two chromoelectric dipoles.  

All the terms in \eqref{E0} are well known in perturbation theory.
The three-loop expression of the color-singlet static potential has been computed in~\cite{Anzai:2009tm,Smirnov:2009fh}.
The Wilson coefficient $V_A$ is 1 up to corrections of order $\als^2$~\cite{Brambilla:2006wp}.
The chromoelectric field correlator, $\langle {\rm Tr}\,\{g{\bf E}(t,{\bf 0}) \cdot g{\bf E}(0,{\bf 0})\}\rangle$ (Wilson lines connecting the fields 
are understood), is known up to two loops~\cite{Eidemuller:1997bb}.
The color-octet static potential is known up to three loops~\cite{Anzai:2013tja}.

An advantage of equation \eqref{E0} is that it allows for an efficient resummation of the logarithmic contributions 
to the static potentials and eventually to the static energy. This is achieved by evaluating the 
anomalous dimensions of the potentials through the computation of the ultraviolet divergences in the relevant integrals 
proportional to the two chromoelectric dipoles and then by solving the renormalization group equations for the static 
potentials~\cite{Pineda:2000gza,Brambilla:2009bi}. This is how contributions to the static energy  
of the type $\als^{3+n}\ln^n\als/r$ and $\als^{4+n}\ln^n\als/r$ have been computed for all $n$.

The perturbative expansion of $V_s$ is affected by a renormalon ambiguity of order $\lQ$.
The ambiguity reflects in the poor convergence of the perturbative series.

A first method to cure the poor convergence of the perturbative series of $V_s$ consists in subtracting a (constant) 
series in $\als$ from $V_s$ and reabsorb it in a redefinition of the normalization constant $\Lambda_s$~\cite{Pineda:2002se}.
This is the strategy we followed in the earlier analysis~\cite{Bazavov:2012ka}.

A second possibility consists in considering the force 
\be
F(r,\nu) = \frac{d}{d r}E_0(r).
\label{force}
\ee
It does not depend on $\Lambda_s$ and is free from the renormalon of order $\lQ$.
In \eqref{force} we have written explicitly the dependence of $F$ on the renormalization scale for the coupling, $\nu$, 
which appears when computing $F$ order by order in $\als(\nu)$.
Once integrated upon the distance, the force gives back the static energy
\be
E_0(r)=\int_{r_*}^{r}dr'\, F(r',1/r'),
\label{E0force}
\ee
up to an irrelevant constant determined by the arbitrary distance $r_*$, 
which can be reabsorbed in the overall normalization when comparing with lattice data.
Notice that the integration in \eqref{E0force} can be done (numerically) keeping the strong-coupling constant running 
at a natural scale of the order of the inverse of the distance. This scheme does not generate 
therefore potentially large logarithms of the type $\ln \nu r$.
Such logarithms are typically generated within the first method.
In the newer analysis of~\cite{Bazavov:2014soa}, we have adopted this second method.

\section{Numerical analysis}
\label{analysis}
In order to determine $\als$, we match the three-loop expression of $E_0(r)$ in the short range 
with the determination of the same quantity in 2+1-flavor lattice QCD obtained from the tree-level improved gauge action 
and the Highly-Improved Staggered Quark (HISQ) action by the HotQCD collaboration~\cite{Bazavov:2014pvz}.
The strange quark mass, $m_s$, has been fixed to its physical value, while the light quark masses, $m_l$, 
have been set equal to $m_s/20$. This corresponds to a pion mass of about $160$ MeV in the continuum limit.
Several lattice spacings and volumes have been considered, but for our determination of $\als$ we restricted 
to the three finest spacings shown in table~\ref{tablattice}.
The largest gauge coupling, $\beta=7.825$, corresponds to a lattice spacing of $a=0.041$~fm.
The lattice spacing has been fixed using the $r_1$ scale defined as $r^2 d E_0(r)/d r|_{r=r_1}=1.0$.
In turn, the scale $r_1$ could be fixed from the pion decay constant~\cite{Bazavov:2010hj}: $r_1=0.3106\pm0.0017$~fm.

\begin{table}
\begin{center}
\begin{tabular}{|c|ccc|}
\hline
$\beta$ & 7.373     & 7.596     & 7.825     \\
\hline
$r_1/a$ & 5.172(34) & 6.336(56) & 7.690(58) \\
\hline
Volume & $48^3\times64$ & $64^4$ & $64^4$\\
\hline
\end{tabular}
\caption{Lattice couplings, spacings and volumes used for the extraction of $\als$.}
\label{tablattice}
\end{center}
\end{table}

\begin{figure}[ht]
\centering
\includegraphics[width=7cm]{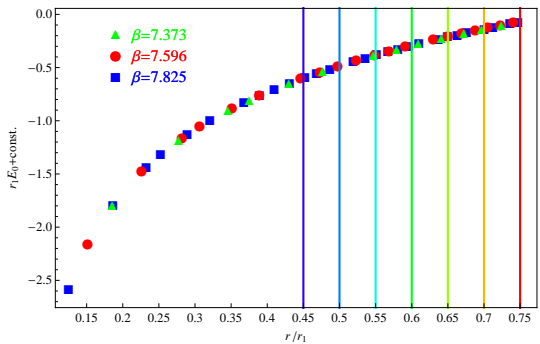}
\caption{Data sets used in the analysis of~\cite{Bazavov:2014soa}. Lattice data are from~~\cite{Bazavov:2014pvz}.}
\label{figdata}
\end{figure}

Because of the many lattice points available at short distance (see figure~\ref{figdata}) we can fit $\als$ (or equivalently $\Lambda_{\MS}$)
for each lattice spacing separately and eventually average these determinations. This is different from our earlier analysis 
in~\cite{Bazavov:2012ka}, where, because of the less data points, we had to fit $\als$ on data points 
from all lattice spacings at the same time. Such a procedure introduced a sizeable extra error source 
due to the different lattice normalizations. This error is absent in the newer analysis in~\cite{Bazavov:2014soa}.

We use the following procedure: {\it (i)} for each lattice spacing we fit $\Lambda_{\MS}$ at different orders of perturbative accuracy;
the overall normalization constant that matches the lattice data with the perturbative expression of $E_0(r)$
is typically fixed on the 7th, 8th or 9th lattice point called $N_{\rm ref}$ in some of the following figures
(this choice is dictated by the fact that lattice points at short distance are expected to be better described by perturbation theory, 
but at too short distance may be affected by lattice artefacts);
{\it (ii)} we repeat the fits for each of the following distance ranges: 
$r<0.75r_1$, $r<0.7r_1$, $r<0.65r_1$, $r<0.6r_1$, $r<0.55r_1$, $r<0.5r_1$, and $r<0.45r_1$, 
and use only the ranges where the reduced $\chi^2$ either decreases or does not increase by more than one unit
or is smaller than~1, when increasing the perturbative order;  
{\it (iii)} to estimate the perturbative uncertainty of the result, we repeat the fits
by varying the renormalization scale in the perturbative expansion from $\nu=1/r$ to $\nu=\sqrt{2}/r$ and $\nu=1/(\sqrt{2}r)$,
or by adding/subtracting a term $\pm 4\als^{n+2}/(3r^2)$ to the $n$-loop expression of the force;  
we take the largest uncertainty.

As expected, perturbative QCD describes better the data in the shortest distance ranges.
In parti\-cu\-lar, criterion {\it (ii)} is fulfilled for data in the range $r < 0.6 r_1$ or shorter; 
to be on the most conservative side, in the final result we will use only fits for $r < 0.5 r_1$.

\begin{figure}[ht]
\centering
\includegraphics[width=7cm]{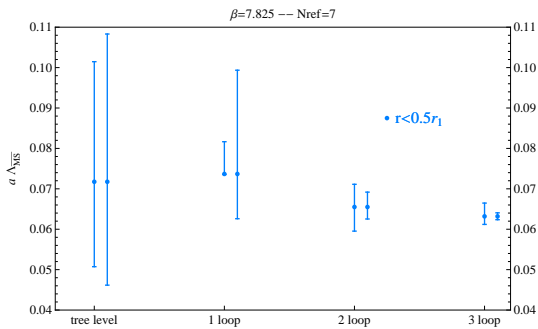}
\caption{Results for $a\Lambda_{\MS}$ from the fits to the lattice data at $\beta = 7.825$ 
for different orders of perturbative accuracy. The first error bar corresponds to the $\nu$ variation,
whereas the second one to the addition of a generic higher-order term. The static energies have been normalized on the 7th point.}
\label{figpert}
\end{figure}

\begin{figure}[ht]
\centering
\includegraphics[width=7cm]{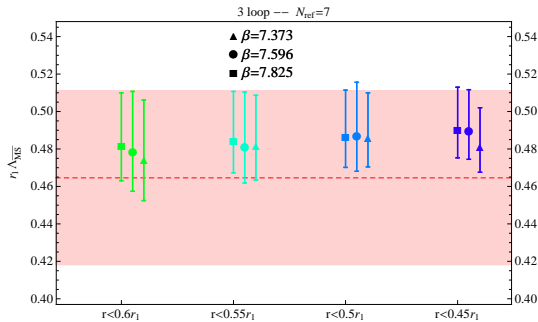}
\caption{Results for $r_1\Lambda_{\MS}$ at three-loop accuracy for the different lattice spacings of table~\ref{tablattice} and ranges of figure~\ref{figdata}. 
We show the fit for data normalized with respect to the 7th lattice point. Similar analyses with consistent results 
have been done in~\cite{Bazavov:2014soa} also for different normalization points.
The band shows our previous result~\cite{Bazavov:2012ka}.}
\label{figrange}
\end{figure}

We have performed several checks and estimated several error sources.
{\it (a)}~When considering the perturbative errors of {\it (iii)} we notice that 
uncertainties due to the higher-order term  $\pm 4\als^{n+2}/(3r^2)$ dominate for the static energy at tree level and one loop, 
while starting from two loops the dominant source of uncertainties comes from variation of $\nu$. Uncertainties decrease in going to higher orders,  
and in all cases they fall inside the error bar of the lower-order determination, see figure~\ref{figpert}.
{\it (b)}~The result is insensitive or very little sensitive to changes in the lattice spacing, in the 
range of short distances used (see figure~\ref{figrange}) and in the point chosen for the normalization with the lattice data.
{\it (c)}~We have estimated and added to our determination of $\Lambda_{\MS}$ a statistical error taken as 
the variation in $\Lambda_{\MS}$ when one allows for the fit to be one $\chi^2$ unit above minimum. 
Statistical errors are smaller than perturbative errors, but, in particular for the shortest distance ranges, not negligible.
{\it (d)}~An analysis done comparing the perturbative expression of the force \eqref{force} with the numerical derivative 
of the lattice data for $E_0(r)$ gives consistent results, but with larger uncertainties associated with the interpolation of the lattice points. 
{\it (e)}~By repeating the fits adding a monomial term proportional to $r^3$ and $r^2$, 
which could be associated with gluon and quark local condensates, and also a term proportional to $r$, 
we could not find evidence for a significant non-perturbative term at short distances and the value of $\Lambda_{\MS}$ remains unchanged.

The analysis revealed that the data are not sensitive to logarithmic corrections of order $\als^{4+n}\ln^n\als/r$, 
moreover these corrections are much smaller than the non-logarithmic contribution of order $\als^4/r$, 
which turns out to be numerically as large as the  logarithmic corrections of order $\als^{3+n}\ln^n\als/r$, but with opposite sign. 
Hence we have included in our perturbative expression of the static energy only logarithmic corrections of the type $\als^{3+n}\ln^n\als/r$, 
whereas we have not included logarithmic corrections of the type $\als^{4+n}\ln^n\als/r$, although they are known.
We chose $\mu = 1.26 r_1^{-1} \sim 0.8$ GeV for the factorization scale.
Variations of $\mu$ only produce small effects on the results.

As shown in figure~\ref{figrange}, the newer analysis of~\cite{Bazavov:2014soa} is consistent with the older 
one of~\cite{Bazavov:2012ka}, but with an error which is roughly half. The reduction in the error 
can be traced back to two main improvements. First, as already discussed in this section, 
the newest lattice data provide much more points in the short range so that we can fit for each lattice spacing independently 
and avoid normalization errors that arise when combining data from different lattices. Second, as discussed at the 
end of section~\ref{theory}, fitting with the perturbative expression obtained from \eqref{E0force} 
allows avoiding uncertainties associated with a fixed renormalization scale $\nu$, which is a potential source of large logarithms of the type $\ln \nu r$.

\section{Results}
\label{results}
The results of the fits of $\Lambda_{\MS}$ for the different lattice spacings 
are summarized in table~\ref{tablambda}.
The first error is the perturbative one, the second error is the one from the statistical uncertainties in the fit, 
and, for the last column, the third error corresponds to the one coming from the conversion from $a$ to $r_1$.
Errors in the right-hand side of the last column are added in quadrature.
The final number for $r_1\Lambda_{\MS}$ with three flavors is obtained as a weighted
average of the results for the three different lattices with errors added linearly.
Once converted in physical units, it reads 
\be
\Lambda_{\MS}=315^{+18}_{-12}~{\rm MeV}.
\label{lambdaMSbar}
\ee

\begin{table}[ht]
\centering
\includegraphics[width=14cm]{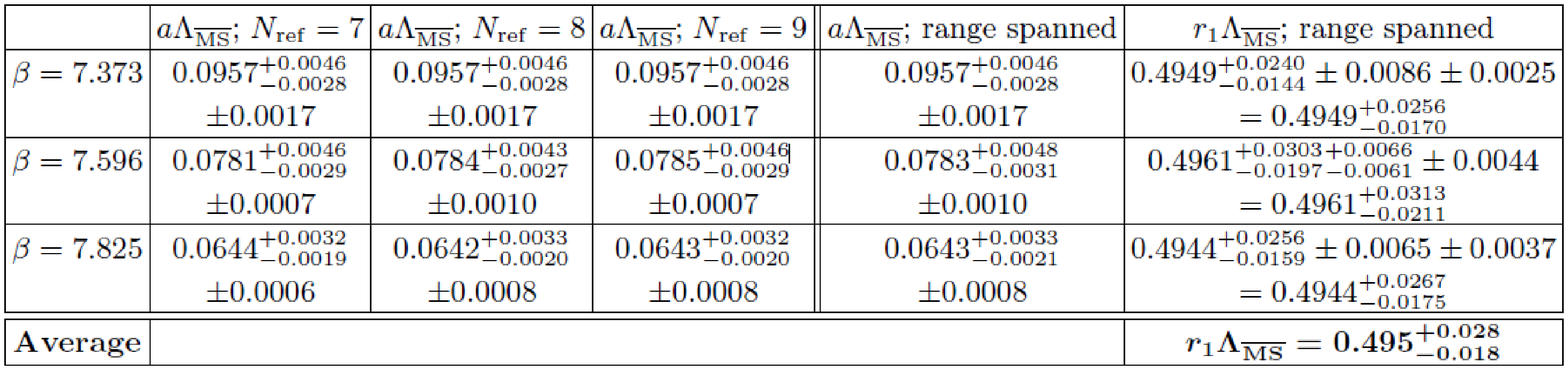}
\caption{$\Lambda_{\MS}$ determinations from the three lattice spacings and averaging.}
\label{tablambda}
\end{table}

\begin{figure}[ht]
\makebox[0cm]{\phantom b}
\put(0,0){\epsfxsize=6.8truecm \epsffile{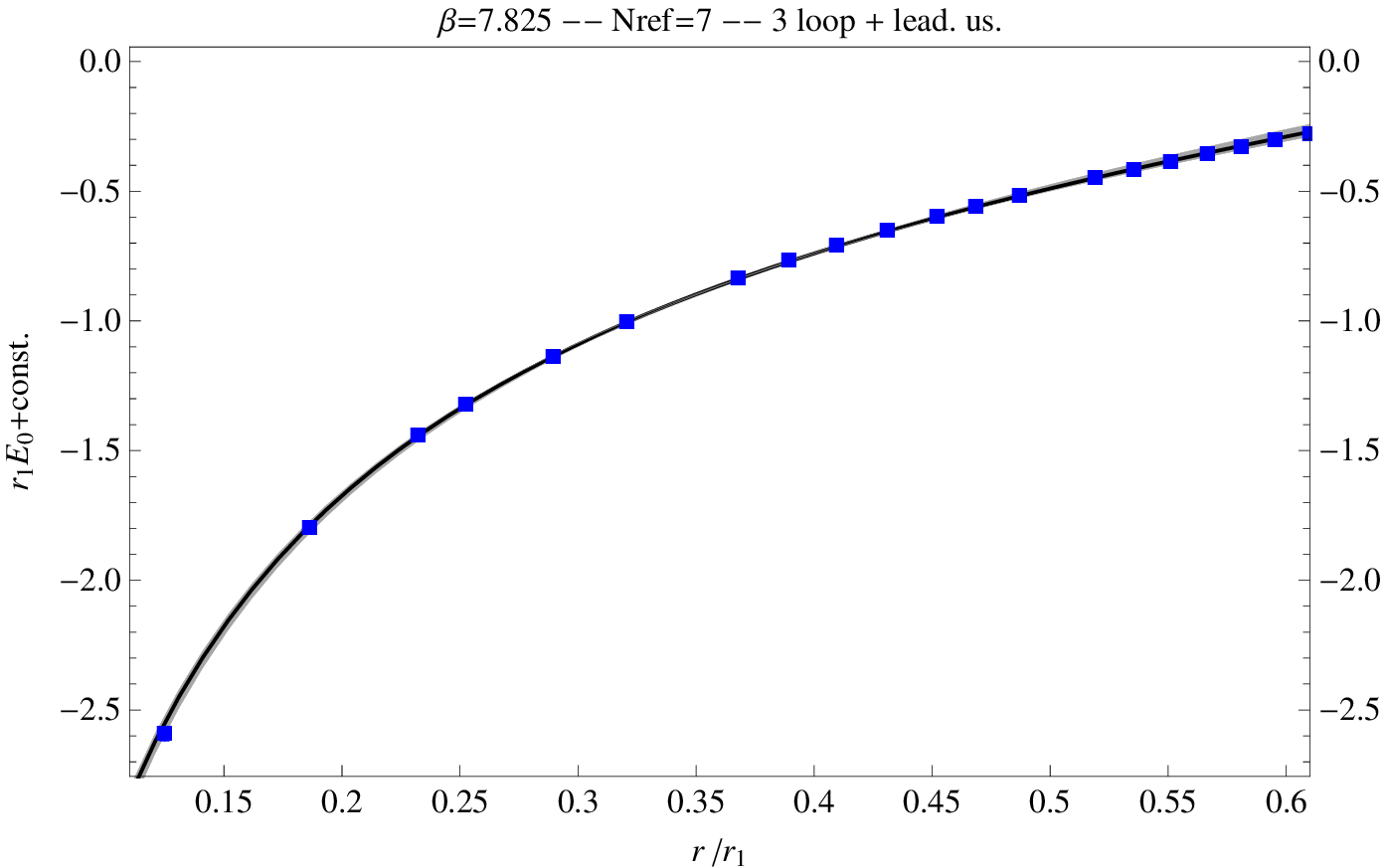}}
\put(200,0){\epsfxsize=6.8truecm \epsffile{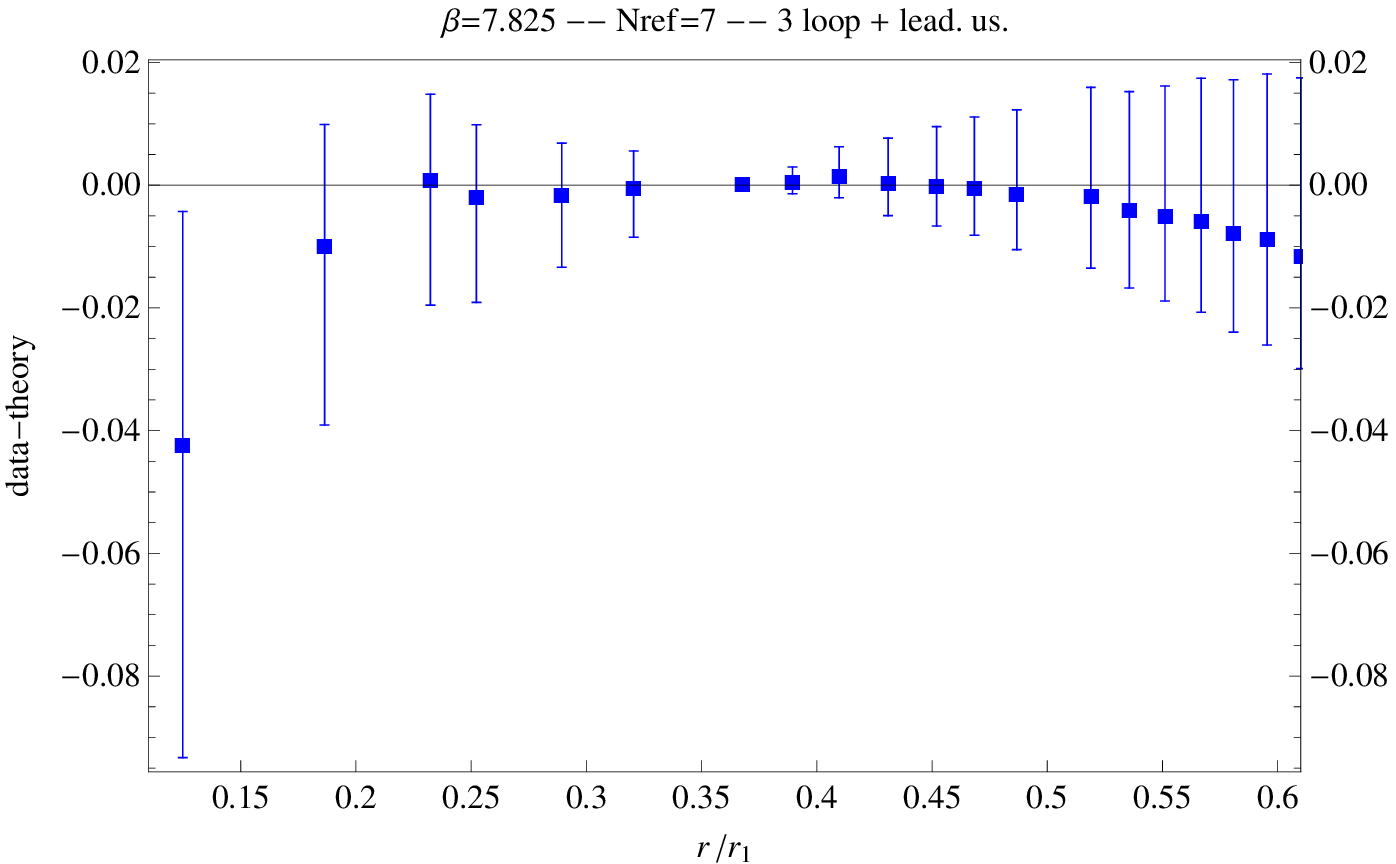}}
\caption{Left panel: comparison of the lattice data for $\beta=7.825$
with the perturbative expression at three loops plus 
logarithmic corrections of the type $\als^{3+n}\ln^n\als/r$. 
We take $r_1\Lambda_{\MS}=0.495^{+0.028}_{-0.018}$; the grey band reflects 
the uncertainty in  $r_1\Lambda_{\MS}$.
Right panel: result of subtracting the perturbative expression from the lattice data.
In both panels the static energy has been normalized on the 7th lattice point.}
\label{figcomparison}
\end{figure}

With this value of $\Lambda_{\MS}$ we can compare the static energy as computed in perturbative QCD
at three loop accuracy with resummation of the $\als^{3+n}\ln^n\als/r$ terms 
with lattice data. In figure~\ref{figcomparison} this comparison is done for the $\beta=7.825$ lattice data.
The comparison shows also visually (the numerical quantitative analysis has been summarized in the previous section) 
that perturbation theory agrees with lattice data up to about 0.2 fm.

By converting \eqref{lambdaMSbar} to $\als$ at the highest energy scale we used, 
i.e., 5.1 GeV, and then evolving the value of $\als$ down to 1.5 GeV with $3$ flavors, 
we obtain 
\be
\als(1.5~{\rm GeV})=0.336^{+0.012}_{-0.008} ,
\label{alphasmc}
\ee
which is one of the most accurate determinations of the strong coupling constant 
at this low-energy scale. Finally, by evolving $\als$ at four loops up to the $Z$ mass, $M_Z$,  
including the decoupling relations at the quark thresholds ($M_c =1.6$~GeV and $M_b=4.7$~GeV), we obtain 
\be
\als(M_Z)=0.1166^{+0.0012}_{-0.0008}.
\label{alphasmZ}
\ee
The effects of higher-order terms in the running are negligible with the current accuracies,
but may become relevant when and if the precision of $\als(M_Z)$ will be reduced to the per mil level.

\section{Outlook}
\label{outlook}
We have computed the strong coupling constant, $\als$, by fitting the static energy, 
as obtained in perturbation theory with three loop accuracy and with resummation of the $\als^{3+n}\ln^n\als/r$ terms,  
with the short-distance part of the static energy as obtained on physical lattices. 
The result is a very accurate determination of $\als$ at low energy. 
At the $Z$-mass scale the determination is competitive with the others entering the PDG average. 
Its central value is lower than the PDG average of 2014~\cite{Agashe:2014kda}, 
but closer to the expected PDG average of 2016~\cite{Bethke2015}. 

There are several ways in which the present determination may be improved with the potential of becoming
the most precise determination of $\als$. As shown in figure~\ref{figpert}, to achieve the present accuracy the inclusion of the 
three-loop result is crucial but, as discussed at the end of section~\ref{analysis},  
not all of the presently available perturbative information has been used. 
More precise lattice data on finer lattices and with more data points at short distances  
could take advantage of it and improve the determination of $\als$. 
Also, it would be important, in order to reduce possible systematic effects, 
to perform the same study on Wilson loops computed on different lattices with different actions.

A possible systematic effect is due to the finite lattice spacing.
A continuum extrapolation would reduce this effect and allow for a precise determination of the force between 
static charges along the same lines developed in~\cite{Necco:2001xg} for the quenched case.
As we mentioned in section~\ref{analysis}, a study of the force with the present lattice data, 
although consistent with the result obtained from the static energy, did not provide a comparable accuracy 
because of finite lattice spacing effects in the calculation of the slope.

Another possibility consists of computing the force directly from the lattice, i.e., not as the slope of the static energy. 
A matrix element that gives the force, $F$, between a static quark located in ${\bf r}$ and a static antiquark located in ${\bf 0}$ is~\cite{Brambilla:2000gk,Pineda:2000sz}
\be
F(r) = - \lim_{ T\to\infty}
\frac{ \left\langle {\rm Tr} \, {\rm P} \,\hat{\bf r} \cdot g{\bf E}(t,{\bf r}) \exp\left\{i g \oint_{r\times T} dz^\mu \, A_\mu\right\} \right\rangle}
{\left\langle {\rm Tr} \, {\rm P} \exp\left\{i g \oint_{r\times T} dz^\mu \, A_\mu\right\} \right\rangle}.
\label{Ewilson}
\ee
The chromoelectric field ${\bf E}(t,{\bf r})$ on the right-hand side is located at the quark line of the Wilson loop.
It would be interesting to compute this matrix element and examine if, besides being an alternative way to compute 
the force, it may also provide a more accurate determination of $\als$.

\begin{acknowledgement}
I thank Alexei Bazavov, Nora Brambilla, Xavier Garcia i Tormo, P\'eter Petreczky and  Joan Soto 
for a long going collaboration on the subject.
This  work is supported  in  part  by  DFG  and  NSFC  (CRC110).
I also acknowledge financial support from the DFG cluster of excellence ``Origin and structure of the universe"
(www.universe-cluster.de).
\end{acknowledgement}

\end{document}